\documentclass{article}
\usepackage{amsfonts}

\usepackage{amsmath}

\input{tcilatex}

\begin{document}

\title{A survey of the ESR model for an objective reinterpretation of
Quantum Mechanics}
\author{Claudio Garola \\
Department of Mathematics and Physics\\
University of Salento\\
73100 Lecce, Italy\\
E-mail: garola@le.infn.it}
\maketitle

\begin{abstract}
Most scholars concerned with the foundations of quantum mechanics (QM) think
that \textit{contextuality} and \textit{nonlocality} (hence \textit{%
nonobjectivity} of physical properties) are unavoidable features of QM which
follow from the mathematical apparatus of QM. Moreover these features are
usually considered as basic in quantum information processing. Nevertheless
they raise still unsolved problems, as the \textit{objectification problem}
in the quantum theory of measurement. The \textit{extended semantic realism }%
(\textit{ESR})\textit{\ model} offers a possible way out from these
difficulties by embedding the mathematical formalism of QM into a broader
mathematical formalism and reinterpreting quantum probabilities as
conditional on detection rather than absolute. The embedding allows to
recover the formal apparatus of QM within the ESR model, and the
reinterpretation of QM allows to construct a noncontextual hidden variables
theory which justifies the assumptions introduced in the ESR model and
proves its objectivity. According to the ESR model both linear and nonlinear
time evolution occur, depending on the physical environment, as in QM. In
addition, the ESR model, though objective, implies modified Bell's
inequalities that do not conflict with QM, supplies different mathematical
representations of proper and improper mixtures, provides a general
framework in which the local interpretations of the GHZ experiment obtained
by other authors are recovered and explained, and supports an interpretation
of quantum logic which avoids the introduction of the problematic notion of
quantum truth.
\end{abstract}

\section{Introduction.}

Since its birth quantum mechanics (QM) proved to be a theory of outstanding
empirical success, but also a source of problems and paradoxes. These mainly
follow from the proposed interpretations of the theory, which multiplied in
time and are still debated. According to Busch \textit{et al}. [1] these
interpretation can be divided in two classes.

(i) \textit{Statistical interpretations}: QM refers to frequencies of
measurements outcomes only. No reference to microscopic objects should enter
its language.

(ii) \textit{Ontic}, or \textit{realistic} interpretations: QM deals with
items of physical systems, or \textit{individual objects}, and their
properties.

The statistical interpretations avoid many problems but can be criticized
from several viewpoints. They imply indeed an instrumentalist view and lack
explanatory power. Moreover, nowadays experimental physicists often claim
that they can deal with individual objects, not only with statistical
ensembles.

On the other side, the realistic interpretations can be reformulated
avoiding ontological commitments if ``individual object'' is considered as a
term of the theoretical language of QM, interpreted (via observational
language) as a click in a preparing device. But in these interpretations,
however reformulated, a crucial problem occurs, at least if one does not
want to go back to a merely statistical interpretation of QM (now in terms
of ensembles of individual objects rather than measurements outcomes): the 
\textit{nonobjectivity} of physical properties in QM, following from
``no-go'' theorems as Bell-Kochen-Specker's, which proves the \textit{%
contextuality} of QM, and Bell's, which proves the \textit{contextuality at
a distance}, or \textit{nonlocality}, of QM. Indeed, nonobjectivity has some
well known intriguing consequences.

(i) \textit{Objectification problem in the quantum theory of measurement}.
If QM is a universal theory, nonobjectivity extends to properties of
macroscopic objects, against everyday evidence. This problem is illustrated
by famous paradoxes, as Schr\"{o}dinger's cat, Wigner's friend, etc.

(ii) \textit{No intuitive model for QM can be provided (wave-particle
duality)}. Every such model would indeed imply objectivity of all properties.

(iii) \textit{Non-epistemic probability}. Quantum probability does not allow
an ignorance interpretation, for the values of nonobjective properties
cannot be assigned independently of a measurement context, that is,
independently of observation. This feature of QM\ implies some
interpretative problems (in particular, proper and improper mixtures have
the same mathematical representation but different physical interpretations).

(iv) \textit{Quantum truth and quantum logic}. The classical notion of truth
as correspondence is unsuitable for the observational language of QM, for no
extension made up of individual objects can be associated with a property
that is nonobjective in a given state of $\Omega $. Hence, a non-classical
notion of truth is required.

Notwithstanding the problematic consequences summarized above, contextuality
and nonlocality are usually maintained to be distinguishing features of QM
whenever one does not explicitly restrict to a statistical interpretation,
independently of the foundational approach that is adopted (e.g., in the
quantum logical, in the operational and in the algebraic approach, in Bohm's
theory, etc.). Moreover, nowdays quantum information theory considers
contextual and nonlocal correlations as basic resources for quantum
information processing and has inspired new foundational approaches, as
Zeilinger's [3], Clifton-Bub-Halvorson's [4], etc. It must be stressed that
the acceptance of contextuality and nonlocality in these approaches stands
not only on the ``no-go'' theorems but also on a series of experimental
results that started with the famous Aspect's experiments [5-7].

Philosophers of science know, however, that no set of experimental results
may determine in a unique way a theory that explains them. Moreover, every
``no-go'' theorem follows from assumptions (some of which are often left
implicit) that can be questioned. Several years ago a research was therefore
started by the author, together with some collaborators, with the aim of
inquiring whether it was possible to recover objectivity by embedding the
mathematical apparatus of QM into a broader mathematical framework and
reinterpreting it in such a way to turn around the ``no-go'' theorems. Of
course, this new framework had to satisfy a basic requirement, that is, it
had to explain the experimental results mentioned above and, more generally,
the empirical success of QM. This research has been recently completed with
the proposal of a new theory called \textit{ESR} (\textit{extended semantic
realism}) \textit{model} [8-17]. The main features of this model are resumed
in Sect. 3, and some results obtained by discussing known problems of QM in
the new framework are presented in Sect. 4. Sect. 2 is instead devoted to a
preliminary clarification of the notion of nonobjectivity.

\section{On the notion of nonobjectivity}

To make the notion of nonobjectivity that will be used in this paper more
precise, let us firstly recall that a physical system $\Omega $ is usually
associated in QM with a set $\mathcal{S}$ of \textit{states} and a set $%
\mathcal{O}$ of \textit{observables}. The set $\mathcal{S}$ is partitioned
into a subset $\mathcal{P}$ of \textit{pure} states and a subset $\mathcal{M}
$ of \textit{mixtures}. Furthermore, a (\textit{physical})\textit{\ property}
is defined as a pair $F=(A,\Sigma )$, with $A\in \mathcal{O}$ and $\Sigma $
a Borel subset of the set $\Xi $ of all possible values of $A$ [1, 2]. The
physical system $\Omega $ can then be characterized by a triple $(\mathcal{S}%
,\mathcal{F},p)$ [1, 18]), where $\mathcal{F}$ is the set of all properties
of\ $\Omega $ and $p$ is a probability function

\begin{center}
$p:(S,F)\in \mathcal{S}\times \mathcal{F}\longrightarrow \lbrack 0,1]$.
\end{center}

Because of the characterization above, properties play a fundamental role in
the foundations of QM. Given a property $F=(A,\Sigma )\in \mathcal{F}$, one
says that $F$ has truth value \textit{true} (\textit{false}) iff the value
of $A$ belongs (does not belong) to $\Sigma $. If one adopts the realistic
interpretation of QM (Sect. 1), every property $F$ is in principle
measurable (but different properties may be not simultaneously measurable)
on an individual object $a$, that is, an item of $\Omega $. The standard \
formulations of QM usually consider only \textit{idealized} (efficiency $1$)
measurements. These measurements are dichotomic and their outcomes are
labeled \textit{yes} and \textit{no}, the former corresponding to the value 
\textit{true} of $F$ and the latter to the value \textit{false}.

The notion of objectivity can now be defined as follows.

\textit{A property }$F$\textit{\ of }$\Omega $\textit{\ is objective for an
individual object }$a$\textit{\ iff its value }(true\textit{/ }false)\textit{%
\ is not only assigned for every measurement context }(value definiteness)%
\textit{\ but also independent of this context.}

It follows from the definition above that a property $F$ is \textit{%
nonobjective} whenever its value is not assigned for every measurement
context or, if assigned, it depends on the context. Based on this definition
one concludes that, if one does not want to reduce to a merely statistical
interpretation, the realistic interpretations of QM imply that QM is a
nonobjective theory, in the sense that, for every individual object $a$ in a
given state $S$ of a physical system $\Omega $, there are both properties
that can be considered objective and properties that must be considered
nonobjective. To be precise, if an individual object $\alpha $\ is in the
state $S$, then $F$ can be considered objective for $\alpha $\ if $p(S,F)\in
\{0,1\}$, but it is necessarily nonobjective if $p(S,F)\notin \{0,1\}$ (note
that this conclusion implies that $F$ is objective for $\alpha $\ if and
only if it is objective for every individual object in the state $S$).

\section{The ESR Model}

The ESR model stems from the intuitive idea that the set of all properties
which are objective for an individual object $\alpha $ in a state $S$ (to be
determined by the model itself, as in QM) may be such that $\alpha $ has
nonzero probability of remaining undetected when a property $F$ is measured
on it. This ``no-detection'' probability may vary with F and $S$ but does
not depend on the device that is used to perform the measurement: hence, it
may be different from $0$ also in the case of exact (efficiency 1)
measurements. The lack of efficiency of real measurements superimposes to
it, usually hiding it.

\subsection{The fundamental assumption}

To formalize the intuitive idea expounded above, the ESR model starts from
the quantum description of a physical system $\Omega $ in terms of states
and observables, but adds a ``no-registration outcome'' $a_{0}$ to the set $%
\Xi $ of all possible values of any quantum observable $A$. The outcome $%
a_{0}$ is considered as a possible result of an exact measurement of $A$ and
not only as the initial position of a pointer that is abandoned when the
measurement is performed. Hence the introduction of $a_{0}$ transforms the
quantum observable $A$ into a \textit{generalized observable} $A_{0}$. This
generalized observable is then associated with a family of properties of the
form $(A_{0},\Sigma )$, where $\Sigma $ is a Borel subset of the set $\Xi
_{0}=\Xi \cup \{a_{0}\}$ of all possible values of $A_{0}$. When $a_{0}$
does not belong to $\Sigma $, the property $F=(A_{0},\Sigma )$ coincides
with the quantum property $(A,\Sigma )$. Therefore the subset $\left\{
(A_{0},\Sigma )\mid a_{0}\notin \Sigma \right\} $ of all properties of this
kind corresponds bijectively to the set of all properties of $\Omega $ in QM
and can be identified with it (hence it is denoted by $\mathcal{F}$ in the
following). Then, the intuitive idea expounded above can be formally
expressed by the \textit{fundamental equation} of the ESR model

\begin{center}
$p^{t}(S,F)=p^{d}(S,F)p(S,F)$.
\end{center}

In this equation $S$ is a state and $F=(A_{0},\Sigma )\in \mathcal{F}$.
Then, $p^{t}(S,F)$ is the overall probability that an idealized measurement
of $F$ performed on an individual object $\alpha $ in the state $S$ yields
outcome \textit{yes}, $p^{d}(S,F)$ is the probability that $\alpha $ is
detected in the measurement (\textit{detection probability}), and $p(S,F)$
is the probability that the measurement yields outcome \textit{yes} when $%
\alpha $ is detected (\textit{conditional on detection probability}).

The \textit{fundamental assumption} of the ESR model can now be stated as
follows.

AX. \textit{Let }$S\in \mathcal{P}$\textit{\ and }$F\in \mathcal{F}$\textit{%
. Then, the probability }$p(S,F)$\textit{\ coincides with the probability
supplied by QM, via Born's rule.}

It is important to note that assumption AX concerns pure states only
(mixtures require indeed a separate treatment, see Sect. 3.3). Furthermore
this assumption has two relevant consequences.

(i) \textit{Conservative}. The ESR model embodies the mathematical formalism
of QM.

(ii) \textit{Innovative}. The ESR model deeply modifies the standard
interpretation of the mathematical formalism of QM. According to QM, Born's
rule supplies an absolute probability (physically interpreted as the large
number limit of the ratio $n/N$, where $n$ is the number of individual
objects in the state $S$ that display the property $F\in \mathcal{F}$ when $%
F $ is measured, and $N$ is the number of individual objects in the state $S$%
). According to the ESR model, if $S$ is pure the same rule supplies a
conditional probability (physically interpreted as the large number limit of
the ratio $n/N^{d}$, where $N^{d}\leq N$ is the number of all individual
objects in the state $S$ that are detected when $F$ is measured).

\subsection{The mathematical representation}

For every $S\in \mathcal{P}$\ and $F=(A_{0},\Sigma )\in \mathcal{F}$, the
introduction of the three probabilities $p^{t}(S,F)$, $p^{d}(S,F)$ and $%
p(S,F)$ in place of the standard quantum probability implies that the
mathematical formalism of QM must be extended to take into account these
probabilities. Such an extension leads to new representations of states,
observables and properties.

\textit{The detection probability} $p^{d}(S,F)$. No theory is available at
present to predict $p^{d}(S,F)$. Hence $p^{d}(S,F)$ is considered as a
parameter in the ESR model, to be determined empirically. It is only
required that $p^{d}(S,F)$ satisfies a mathematical assumption (see below)
that seems quite natural from an intuitive point of view.

\textit{The conditional on detection probability} $p(S,F)$.\ Assumption AX
implies that this probability can be obtained by using standard quantum
rules. Hence, as far as $p(S,F)$ is concerned, the physical system $\Omega $
can be associated with a Hilbert space $\mathcal{H}$. Moreover, a pure state 
$S$ can be represented by a one-dimensional orthogonal projection operator $%
\rho _{S}$ on $\mathcal{H}$, the generalized observable $A_{0}$ can be
represented by the same self-adjoint operator $\hat{A}$ that represents the
observable $A$ of QM from which $A_{0}$ is obtained, and the property $F$
can be represented by an orthogonal projection operator $P^{\hat{A}}(\Sigma
) $ on H. Furthermore, the standard quantum equation holds

\begin{center}
$p(S,F)=Tr[\rho _{S}P^{\hat{A}}(\Sigma )]$.
\end{center}

\textit{The overall probability} $p^{t}(S,F)$. Bearing in mind the
fundamental equation of the ESR model and the mathematical representation of 
$p_{S}(F)$, one obtains

\begin{center}
$p^{t}(S,F)=Tr[p^{d}(S,F)\rho _{S}P^{\hat{A}}(\Sigma )]$.
\end{center}

Hence

\begin{center}
$p^{t}(S,F)=Tr[\rho _{S}T_{S,A_{0}}(\Sigma )]$,
\end{center}

where $T_{S,A_{0}}(\Sigma )=p^{d}(S,F)P^{\hat{A}}(\Sigma )$ is a positive
operator bounded by $0$ and $1$ (\textit{effect}). One then assumes that a
mapping $p_{S,A_{0}}^{d}(\lambda )$ of the set $\Xi $ of all possible values
of $A$ into $[0,1]$ exists such that

\begin{center}
$T_{S,A_{0}}(\Sigma )=\int_{\Sigma }p_{S,A_{0}}^{d}(\lambda )P^{\hat{A}%
}(d\lambda )$
\end{center}

(the existence of $p_{S,A_{0}}^{d}(\lambda )$ constitutes the only
mathematical assumption on $p^{d}(S,F)$ in the ESR model).

The above equations imply that, as far as $p^{t}(S,F)$ is concerned, the
pure state $S$ can still be represented by $\rho _{S}$ and the property $%
(A_{0},\Sigma )$ is represented by a family $\left\{ T_{S,A_{0}}(\Sigma
)\right\} _{S\in \mathcal{P}}$ of effects. Moreover, the generalized
observable $A_{0}$\ is represented by the family of \textit{commutative
operator valued measures}

\begin{center}
$\mathcal{T}_{A_{0}}=\left\{ T_{S,A_{0}}:\Sigma \in \QTR{sl}{B}(\Xi
)\longrightarrow T_{S,A_{0}}(\Sigma )\in \mathfrak{B}(\mathcal{H})\right\}
_{S\in \mathcal{P}}$
\end{center}

where $\QTR{sl}{B}(\Xi )$ is the set of all Borel sets on $\Xi $ and $%
\mathfrak{B}(\mathcal{H})$ the set of all bounded positive operators on $%
\mathcal{H}$.

Putting together the representations of properties to be used in order to
evaluate the conditional on detection probability $p_{S}(F)$ and the overall
probability $p_{S}^{t}(F)$ in the case of pure states, one obtains that a
complete mathematical representation of a property $F=(A_{0},\Sigma )\in 
\mathcal{F}$ is provided in the ESR model by the pair

\begin{center}
$(P^{\hat{A}}(\Sigma ),\left\{ T_{S,A_{0}}(\Sigma )\right\} _{S\in \mathcal{P%
}})$.
\end{center}

Analogously, a complete representation of the generalized observable $A_{0}$
is provided by the pair

\begin{center}
$(\hat{A},\mathcal{T}_{A_{0}})=(\hat{A},\left\{ T_{S,A_{0}}:\Sigma \in 
\QTR{sl}{B}(\Xi )\longrightarrow T_{S,A_{0}}(\Sigma )\in \mathfrak{B}(%
\mathcal{H})\right\} _{S\in \mathcal{P}})$.
\end{center}

The following remarks are then important.

(i) In the representation of $F$\ the first element of the pair coincides
with the standard representation of $F$\ in QM. In the representation of $%
A_{0}$\ the first element of the pair coincides with the standard
representation of the quantum observable $A$ from which $A_{0}$\ is obtained.

(ii) In both representations the second element is a family, parametrized by
the set of pure states. Hence, as far as $p^{t}(S,F)$ is concerned, the
representation of a property, or of an observable, is not given once for
all, because it depends on the state of the individual object on which the
property, or the observable, is measured.

\subsection{Proper and improper mixtures}

The results expounded in Sect.3.2 show that pure states can be represented
in the ESR model by the same density operators that represent them in QM.
One can then wonder whether similar results hold in the case of mixtures.

According to many authors [1, 18, 19] there are in QM \textit{proper} and 
\textit{improper} mixtures, which are mathematically represented in the same
way (density operators) but have different operational definitions, which
imply different interpretations of the coefficients that occur in their
decompositions in terms of pure states (epistemic versus nonepistemic
probabilities).

In the ESR model these two kinds of mixtures have different mathematical
representations, corresponding to their different operational definitions
[10, 12, 13, 17], as follows.

(i) \textit{Improper mixtures}. These mixtures can be represented by the
same density operators that represent them in QM. Assumption AX can be
extended to improper mixtures by substituting the subset $\mathcal{P}$ of
all pure states with the subset $\mathcal{P}\cup \mathcal{N}$, where $%
\mathcal{N}$ is the subset of all improper mixtures. The representations of
properties and observables can then be extended to improper mixtures by
introducing the same substitution. Hence improper mixtures are considered as 
\textit{generalized pure states} in the ESR model.

(ii) \textit{Proper mixtures}. Each proper mixture has a rather complicated
representation as a family of pairs parametrized by the set $\mathcal{F}$ of
properties. Each pair in the family consists of a density operator and a
detection probability. The explicit form of these mathematical entities is
given in [10, 12, 13] and will not be reported here for the sake of brevity.

\subsection{The generalized L\"{u}ders postulate}

In QM the L\"{u}ders postulate selects a subset of (exact) \textit{ideal
first kind measurements} that change a state according to a prefixeed rule
[2]. This postulate is generalized in the ESR model as follows.

Consider an exact dichotomic measurement $\mathfrak{M}$\ of a property $%
F=(A_{0},\Sigma )\in \mathcal{F}$ on an individual object $\alpha $ in the
state $S$, with $S$ a pure state or an improper mixture. Then $\mathfrak{M}$
is an \textit{idealized} measurement of $F$ if the state $S_{F}$ after the
measurement is represented by the density operator

\begin{center}
$\rho _{S_{F}}=\frac{T_{S,A_{0}}(\Sigma )\rho _{S}T_{S,A_{0}}^{\dagger
}(\Sigma )}{Tr\left[ T_{S,A_{0}}(\Sigma )\rho _{S}T_{S,A_{0}}^{\dagger
}(\Sigma )\right] }$
\end{center}

whenever the \textit{yes} outcome is obtained,.

By analogy with QM, the rule expressed by the equation above is called 
\textit{the generalized L\"{u}ders postulate}. It must be stressed that it
does not apply directly to proper mixtures (which are not represented as in
QM, see Sect. 3.3). However, the representation of the final state in the
case of proper mixtures can be deduced from the equation above. Its
mathematical form is rather complicated [10, 12, 13] and will not be
reported here for the sake of simplicity.

\subsection{Time evolution}

The ESR model has been recently completed by studying time evolution [17],
based on the idea that the generalized L\"{u}ders postulate supplies an
example of the change of state of an individual object interacting with
another object (the measuring apparatus). Indeed this example provides some
suggestions for the dynamics of the composite system of the two objects. In
particular, a crucial difference from time evolution in QM occurs because
the generalized L\"{u}ders postulate introduces a change of state also in
the case of individual objects that are not detected by the measurement.

The following conclusions are attained in the case of pure states or
improper mixtures (the details of the treatment will not be reported here
for the sake of brevity).

(i) One can assume that closed systems undergo linear Hamiltonian evolution,
as in QM.

(ii) The evolution of open systems may be linear or not, depending on their
interaction with the environment, as in QM.

(iii) The evolution induced by a measurement on an individual object is
necessarily nonlinear.

The results above show that time evolution in the ESR model matches time
evolution in QM, but for the distinguishing feature in item (iii). One can
then prove that time evolution in the case of proper mixtures can be deduced
from time evolution in the case of pure and generalized pure states if an
obvious assumption is introduced.

\subsection{H.V. models and objectivity}

It remains to discuss the crucial issue of nonobjectivity. Indeed, the main
aim of the ESR model is supplying an objective theory, embodying from one
side the basic mathematical formalism of QM and avoiding, on the other side,
the problems following from nonobjectivity (Sect. 1).

The proof of the objectivity of the ESR model is obtained by showing that
this model admits noncontextual (hence local) \textit{hidden variables} (%
\textit{h.v.})\textit{\ models} (at variance with earlier formulations
[8-16], the latest version of the ESR model [17] does not introduce h.v.
from the beginning). To this end a set $\mathcal{F}_{\mu }$\ of \textit{%
microscopic properties} of the physical system $\Omega $ is introduced which
is in one-to-one correspondence with the set $\mathcal{F}$ of (macroscopic)
properties. For every individual object $\alpha $, the set $\mathcal{F}_{\mu
}$ is then partitioned into two subsets, the subset $s$ of all the
microscopic properties that are \textit{possessed} by $\alpha $\ and the
subset $\mathcal{F}_{\mu }\setminus s$ of all the microscopic properties
that are \textit{not possessed} by $\alpha $. The subset $s$ is called 
\textit{the microscopic state} of $\alpha $. Then, new overall probability,
detection probability and conditional on detection probability are
introduced referring to the microscopic state $s$ of $\alpha $ rather than
to its (macroscopic) state $S$. By introducing the further probability $%
p(S\mid s)$ that an individual object $\alpha $ in the state $S$ is in the
microscopic state $s$, one can deduce the fundamental equation of the ESR
model, thus obtaining the desired noncontextual h.v. model.

Because of the above result and of the one-to-one correspondence between $%
\mathcal{F}_{\mu }$\ and $\mathcal{F}$, one concludes that all properties in 
$\mathcal{F}$ can be considered objective in the sense specified in Sect. 1.
Hence the ESR model is an objective theory. It follows in particular that
quantum probabilities can be considered epistemic, so that no
objectification problem occurs. Of course, this result finds its roots in
the reinterpretation of quantum probabilities as conditional on detection
rather than absolute (Sect. 3.1), which allows to turn around the ``no-go''
theorems of QM (Sect. 4.1).

\subsection{Empirical consequences}

As we have anticipated in Sect. 1, the empirical success of QM imposes a
fundamental constraint on every attempt at modifying QM to avoid the
problems following from nonobjectivity. The predictions of QM that have been
experimentally verified must in fact be reproduced by the new theory within
the limits of the experimental errors. On the other side, the new theory
should also provide some testable predictions that make it empirically
different from QM, allowing one to check which theory is correct. The ESR
model satisfies both these conditions. Indeed, the predictions of the ESR
model in experiments on overall probabilities are formally different from
the predictions of QM, but, if the state $S$ of the individual objects that
are considered is a pure state or an improper mixture, they may be close to
the quantum predictions whenever the detection probabilities are close to $1$%
. Moreover the predictions of the ESR model in experiments on conditional on
detection probabilities (as Aspect's experiments, in which non-detected
individual objects are not taken into account [5, 6]) are identical to the
predictions of QM. The predictions of the ESR model in experiments on
overall probabilities in which the state $S$ of the individual objects that
are considered is a proper mixture may be instead very different from the
predictions of QM and single out a class of experiments that can distinguish
the two theories.

\section{Applications}

The ESR model has been used to deal with some well known problematical
issues in QM. The obtained results can be resumed as follows.

\subsection{The ``no-go'' theorems}

Because of Assumption AX, the ``no-go'' theorems of QM do not hold in the
ESR model [9, 11, 16]. This relevant result can be intuitively explained as
follows.

By considering only the Bell-KS and Bell theorems for the sake of brevity,
one sees that all proofs that do not resort to inequalities proceed ab
absurdo. They consider some different quantum laws linking together physical
properties of an individual object $\alpha $ and show that a contradiction
occurs if all properties of $\alpha $ are supposed to be objective. The laws
that are chosen, however, cannot be checked simultaneously. Indeed each of
them contains some observables that are incompatible with some of the
observables that occur in the other laws. Suppose that a measurement is
performed on $\alpha $ to check one of the laws, and that $\alpha $ is
detected. Then, the law will be confirmed. But one cannot simultaneously
check some of the remaining laws, and cannot exclude that the objective
properties of $\alpha $ be such that $\alpha $ would not be detected if such
a check were done. Thus, the assumption that all laws must simultaneously
hold for $\alpha $ is arbitrary in the framework of the ESR model, and no
``conspiracy of nature'' is required to reach this conclusion. Hence the
aforesaid proofs of contextuality and nonlocality, that imply
nonobjectivity, rest on a questionable assumption from the point of view of
the ESR model.

A similar line of argument holds when considering the proofs of the Bell
theorem that resort to inequalities. Indeed Bell's inequalities do not hold
in the ESR model at a macroscopic level, notwithstanding nonobjectivity
(they hold instead in the h.v. models discussed in Sect. 3.6, at a purely
theoretical microscopic level). When considering, for instance, the original
Bell's inequality, one obtains that it must be replaced by the equation

\begin{center}
$\mid E(A_{0}(a),B_{0}(b))-E(A_{0}(a),B_{0}(c))\mid \leq
1+E(A_{0}(b),B_{0}(c))$
\end{center}

where $E(A_{0}(a),B_{0}(b))$ denotes the expectation value of the products
of the \textit{trichotomic} observables $A_{0}(a)$ and $B_{0}(b)$, depending
on the parameters $a$ and $b$, respectively. The symbols $%
E(A_{0}(a),B_{0}(c))$ and $E(A_{0}(b),B_{0}(c))$\ have similar meanings.

Analogously, the Clauser-Horne--Shimony-Holt inequality must be replaced by
the equation

\begin{center}
$\mid E(A_{0}(a),B_{0}(b))-E(A_{0}(a),B_{0}(c))\mid +\mid
E(A_{0}(d),B_{0}(b))-E(A_{0}(d),B_{0}(c))\mid $

$\leq 2$.
\end{center}

These modified Bell's inequalities, do not necessarily contrast with quantum
inequalities. Hence also the proofs of nonlocality resorting to inequalities
are invalid in the ESR model.

Rather than objectivity, the ESR model questions the unrestricted validity
of quantum laws. Assumption AX implies indeed that a quantum law holds for
an individual objects $\alpha $ if $\alpha $ is detected when the law is
checked on it, while it does not necessarily hold if $\alpha $ remains
undetected because of its objective properties (that are not uniquely
determined by the state $S$ of $\alpha $ in the ESR model).

\subsection{The GHZ experiment}

The general h.v. models for the ESR model can be used to produce h.v. models
for specific physical situations and experiments.

In particular, it has been recently proved that the finite ``toy models''
contrived by Szab\'{o} and Fine in 2002 to provide a local explanation of
the Greenberger-Horne-Zeilinger (GHZ) experiment can be obtained as special
cases of the foregoing general h.v. models [14].

\subsection{Quantum logic and quantum truth}

It has also been recently shown that quantum logic can be embedded into a
suitable extended classical logic, the embedding preserving the logical
order but not the algebraic structure [15].

The above result must be considered as purely formal if one accepts the
standard interpretation of QM. It acquires instead a physical interpretation
in the ESR model because of objectivity of properties in this model.
Objectivity indeed allows one to consider the set of individual objects
formally associated with every $F\in \mathcal{F}$ as the set of all objects
that \textit{possess} the property $F$. It follows that no notion of quantum
truth, different from classical truth and incompatible with it is needed in
the ESR model. Rather, quantum logic can be seen as a mathematical structure
formalizing the metalinguistic notion of \textit{verifiability} according to
QM.


\begin{thebibliography}{10}
\bibitem[1]{BLM96} Busch, P., Lahti, P.J., Mittelstaedt, P.: The Quantum
Theory of Measurement. Springer, Berlin (1991, 1996).

\bibitem[2]{BandC81} Beltrametti, E.G., Cassinelli, G.: The Logic of Quantum
Mechanics. Addison-Wesley, Reading (1981).

\bibitem[3]{Zei} Zeilinger, A.: A foundational principle for quantum
mechanics. Found. Phys. \textbf{29}, 631-643 (1999).

\bibitem[4]{Clift} Clifton, R., Bub, J. Halvorson, H.: Characterizing
quantum theory in terms of informaation theoretic constraints. Found. Phys.%
\textbf{33}, 1561-1591 (2003).

\bibitem[5]{Asp.a} Aspect, A., Grangier, P., Roger, G.: Experimental
realization of Einstein-Podolski-Rosen-Bohm gedankenexperiment: A new
violation of Bell's inequalities. Phys. Rev. Lett. \textbf{49}, 91-94 (1982).

\bibitem[6]{Asp.b} Aspect, A., Dalibard, J., Roger, G.,: Experimental test
of Bell's inequalities using time-varying analyzers. Phys. Rev. Lett. 
\textbf{49}, 1804-1807 (1982).

\bibitem[7]{Gen} Genovese, M.: Research on hidden variables theories: A
review on recent progresses. Phys. Repts. \textbf{413}, 319-396 (2005).

\bibitem[8]{GS2009} Garola, C., Sozzo, S.: The ESR model: a proposal for a
noncontextual and local Hilbert space extensions of QM. Europhys. Lett. 
\textbf{86}, 20009--20015 (2009).

\bibitem[9]{GS2010} Garola, C., Sozzo, S.: Embedding quantum mechanics into
a broader noncontextual theory: a conciliatory result. Int. J. Theor. Phys.
49, 3101--3117 (2010).

\bibitem[10]{GS2011a} Garola, C., Sozzo, S.: Generalized observables, Bell's
inequalities and mixtures in the ESR model. Found. Phys. \textbf{41},
424--449 (2011).

\bibitem[11]{GS2011b} Garola, C., Sozzo, S.: The modified Bell inequality
and its physical implications in the ESR model. Int. J. Theor. Phys. \textbf{%
50}, 3787--3799 (2011).

\bibitem[12]{GS2011c} Garola, C., Sozzo, S.: Representation and
interpretation of mixtures in the ESR model. Theor. Math. Phys. \textbf{168}%
, 912--923 (2011).

\bibitem[13]{GS2012} Garola, C., Sozzo, S. Extended representations of
observables and states for a noncontextual reinterpretation of QM. J. Phys.
A: Math. Theor. \textbf{45}, 075303--075315 (2012).

\bibitem[14]{GPPS} Garola, C., Persano, M., Pykacz, J., Sozzo, S.: Finite
local models for the GHZ experiment. Int. J. Theor. Phys. \textbf{53},
622-644 (2014).

\bibitem[15]{GS13} Garola, C., Sozzo, S.: Recovering quantum logic within an
extended classical framework. Erkenn. \textbf{78}, 399--419 (2013).

\bibitem[16]{GP13} Garola, C., Persano, M.: Embedding quantum mechanics into
a broader noncontextual theory. Found. Sci., DOI
10.1007/s10699-013-9341-z.(2013).

\bibitem[17]{GSW14} Garola, C., Sozzo, S., Wu, J.: Outline of a
generalization and reinterpretation of quantum mechanics recovering
obectivity. ArXiv: 1402.4394v1 [quant-ph] (2014).

\bibitem[18]{Aer} Aerts, D.: Foundations of quantum physics: A general
realistic and operational approach. Int. J. Theor. Phys. \textbf{38},
289-358 (1999).

\bibitem[19]{D'Esp} d'Espagnat, B.: Conceptual Foundations of Quantum
Mechanics. Benjamin, Reading (1976).

\bibitem[20]{Timp} Timpson, C.G., Brown, H.R.: Proper and improper
separability. Int. J. Quant. Inf. \textbf{3}, 679-690 (2005).
\end{thebibliography}
\end{document}